\documentclass[english]{lni}
\usepackage{bm}
\usepackage{bbm}
\usepackage{physics}

\usepackage{algorithm}
\usepackage{algpseudocode}
\usepackage{listings}

\usepackage{booktabs}
\usepackage{makecell}
\usepackage[table]{xcolor}
\usepackage{subcaption}
\usepackage{tikz}

\usepackage{textcomp}
\usepackage{soul}
\usepackage{stfloats}
\usepackage{balance}
\usepackage{pifont}
\usepackage{xurl}

\usepackage{graphicx}

\begin{document}
%%% Mehrere Autoren werden durch \and voneinander getrennt.
%%% Die Fußnote enthält die Adresse sowie eine E-Mail-Adresse.
%%% Das optionale Argument (sofern angegeben) wird für die Kopfzeile verwendet.
\title[Ein Kurztitel]{Quantum-Classical Framework for the Joint Optimization of Bidirectional EV Charging and Trip Assignment}
%% \subtitle{Untertitel / Subtitle} % if needed
 \author[1]{Mustafa Musab}{mustafa.musab@iosb-ast.fraunhofer.de}
 {0009-0002-2964-6529}
 \author[2]{Sebastian Schmitt}{sebastian.schmitt@honda-ri.de}{0000-0001-7130-5483}
 \author[3]{Markus Wappler}{markus.wappler@iosb-ast.fraunhofer.de}{}
 \author[3, 4]{Jörg Lässig}{jlaessig@hszg.de}{0000-0002-3169-3802}
 \author[3]{Daniel Müssig}{daniel.muessig@iosb-ast.fraunhofer.de}{0000-0002-2345-1475}
 \affil[1]{Fraunhofer IOSB-AST \\Am Vogelherd 90\\98693 Ilmenau\\Germany}
 \affil[2]{Honda Research Institute Europe GmbH\\ Carl-Legien-Straße 30\\ 63073 Offenbach/Main\\ Germany}
 \affil[3]{Fraunhofer IOSB-AST\\Wilhelmsplatz 11\\02826 Görlitz\\Germany}
 \affil[4]{University of applied Sciences Zittau/Görlitz\\Wilhelmsplatz 11\\02826 Görlitz\\Germany}
\maketitle

% \begin{abstract}
% \begin{abstract}
\begin{abstract}
The integration of bidirectional (V2G) charging allows electric vehicle (EV) fleets to operate as distributed energy assets. However, the joint optimization of charging schedules and trip assignments is NP-hard. Moreover, bidirectional power flows create non-monotonic state-of-charge (SoC) dynamics that weaken standard dual-based Benders cuts. We propose a hybrid quantum-classical Benders decomposition framework. To overcome the qubit overhead associated with encoding feasibility cuts, we introduce logic-based conflict cuts, which require near-zero auxiliary qubits. Furthermore, we develop a sparse recursive SoC formulation that reduces the constraint-matrix density by a factor of $\sim T/3$ in the horizon length $T$.
We demonstrate that our framework recovers MILP-optimal solutions on benchmark instances with a low auxiliary-qubit budget, and, as a minimal proof of concept, we execute a small instance end-to-end on IBM quantum hardware.
\end{abstract}

%%% Beginn des Artikeltexts
% \section{Introduction}
% \section{Introduction}
\section{Introduction and Literature Review}
The accelerating adoption of electric vehicles is reshaping both the
transportation and energy sectors.
Global EV sales exceeded 17 million in 2024, accounting for more than 20\% of all car sales worldwide, and projections 
indicate continued growth~\cite{iea2025}.
This transition presents a dual opportunity, as EVs can contribute to reducing CO$_2$ emissions while also serving as flexible energy storage assets for power systems through bidirectional charging technologies such as Vehicle-to-Grid (V2G) and Vehicle-to-Home (V2H)~\cite{kumar2025comprehensive}.

Driven by this commercial potential, a growing body of literature has explored different formulations of the EV problem, ranging from trip scheduling to bidirectional energy dispatch. Due to the inherent complexity of these models, decomposition-based methods have become the standard tool for tackling large-scale EV optimization. Benders decomposition, in particular, has been applied to EV
charging infrastructure planning~\cite{golsefidi2025,arslan2016}, optimizing car-sharing station location~\cite{ccalik2019benders},
and V2V routing problems~\cite{ren2025electric}.

Quantum computing offers a different paradigm for the combinatorial aspect of this problem. For instance, quantum algorithms such as the Quantum Approximate Optimization Algorithm (QAOA)~\cite{farhi2014} and quantum annealing~\cite{quantum_cslp2024} can be utilized to explore discrete solution spaces. Previous work from our group studied the bidirectional EV charging problem and successfully benchmarked QAOA on small instances~\cite{mussig2023solving}. Recognizing the large qubit cost required for encoding integer variables, our subsequent research investigated $d$-level qudit systems to encode the EV charging problem more efficiently~\cite{deller2023quantum,ekstrom2026qudit}. In the current work, we extend this foundation to the joint optimization of charging planning and trip scheduling for realistic large scale systems.

Despite the wide range of EV use cases studied in the literature, bidirectional charging remains comparatively underexplored. While bidirectional V2G may appear to be a straightforward extension of the unidirectional case, it fundamentally alters the structure of the feasible region and introduces non-monotonic SoC behavior~\footnote{In unidirectional charging, the SoC monotonically increases during idle periods, allowing solvers to simplify the model by evaluating battery limits solely at trip boundaries. Bidirectional discharging breaks this assumption, requiring dense constraints at every timestep.} that complicates the optimization process. To the best of our knowledge, this is the first hybrid quantum-classical framework applied to the joint optimization of bidirectional EV charging and fleet scheduling.
Table~\ref{tab:literature_classification} outlines the studies most closely related to our study.
\definecolor{rowblue}{HTML}{EBF3FA}
\newcommand{\yes}{\ding{51}}
\newcommand{\no}{\textendash}

\begin{table*}[t]
    \centering
    \renewcommand{\arraystretch}{1.4}
    \caption{Classification of relevant literature by problem characteristics and computing paradigm.}
    \label{tab:literature_classification}
    \begin{tabular}{@{}llcccc@{}}
        \toprule
        \textbf{Reference}
            & \textbf{Type}
            & \textbf{Trip assign.}
            & \textbf{Bidir.\ V2G}
            & \textbf{Paradigm}
            & \textbf{Decomposition} \\
        \midrule
        Veshchezerova et al.~\cite{veshchezerova2023hybrid} & EVCA & \yes & \no & Hybrid & Graph \\
        Deller et al.~\cite{deller2023quantum} & EVC & \no & \yes & Quantum & \no \\ 
        Varga et al.~\cite{varga2022}
            & EVSA & \yes & \no & Classical  & Benders    \\
        M\"{u}ssig et al.~\cite{mussig2023solving}
            & EVC & \no & \yes & Quantum    & \no        \\
        Du et al.~\cite{du2025smart}
            & EVOP & \yes & \yes & Classical  & \no        \\
        Ekstr\o m et al.~\cite{ekstrom2026qudit}
            & EVCA & \yes & \yes & Quantum  & \no        \\
        \midrule
        \rowcolor{rowblue}
        \textbf{Our work}
            & \textbf{EVCA} & \yes & \yes & \textbf{Hybrid} & \textbf{Benders} \\
        \bottomrule
    \end{tabular}

    \smallskip
    \raggedright\small
    \textbf{Type:} EVC\,=\,EV charging; EVCA\,=\,EV charging \& trip assignment; EVSA\,=\,EV fleet scheduling \& assignment;
    EVOP\,=\,EV charging \& orienteering problem.\quad
    \textbf{Indicators:} \yes\;yes;\
    \textendash\;not applicable.
\end{table*}

This paper makes the following contributions:

\begin{enumerate}
    \item \textbf{Logic-Based Hybrid Framework:} We propose a hybrid architecture for bidirectional EV charging and trip assignment using logic-based conflict cuts that admit a near-zero-qubit QUBO encoding.
    
    \item \textbf{Recursive SoC Formulation:} We formulate the SoC constraints in a recursive way, resulting in sparse constraint-matrix density.
\end{enumerate}
The remainder of the paper is structured as follows. Section~\ref{sec:problem} describes the scenario considered and the mathematical formulation of the problem. Section~\ref{sec:hybrid} details the development of our hybrid approach. Finally, Section~\ref{sec:results} presents the experimental results and concludes.

% sec.1 Problem Description
% sec.1 Problem Description
\section{Problem Description} \label{sec:problem}
We consider a bidirectional EV charging and trip assignment scenario where an EV fleet operator allocates vehicles to customer requests (trips) while optimizing bidirectional energy dispatch during idle periods. Table~\ref{tab:notation} summarizes the key parameters and notations used in our mathematical formulation.

\begin{table}[htbp]
    \centering
    \renewcommand{\arraystretch}{1.2}
    \caption{Notation and Problem Parameters}
    \label{tab:notation}
    \resizebox{\textwidth}{!}{%
    \begin{tabular}{@{}ll@{\hspace{8mm}}ll@{}}
        \toprule
        \textbf{Symbol} & \textbf{Description} & \textbf{Symbol} & \textbf{Description} \\
        \midrule
        $N_{\mathrm{EV}}$ & Number of EVs, $n \in \{ 1,\dots,N_{\mathrm{EV}} \}$ & $E^0_n$ & Initial SoC of EV $n$ (at $t = 0$) \\
        $T$ & Number of time steps, $t \in \{1,\dots,T \}$ & $E^{\mathrm{cap}}_n$ & Battery capacity of EV $n$ \\
        $R$ & Number of trip requests, $i \in \{1,\dots,R\} $ & $E^{\mathrm{min}}_n$ & Required final SoC of EV $n$ (at $t = T$) \\
        $\Delta t$ & Time step duration (hours) & $E^{\mathrm{trip}}_i$ & Energy required for serving trip $i$ \\
        $c_t$ & Electricity price at time $t$ & $\mathcal{T}_i$ = $[t_i^{\mathrm{start}}, t_i^{\mathrm{end}}]$ & Time window of trip $i$ \\
        $l_{\mathrm{max}}$ & Maximum charging/discharging power & $P^{\mathrm{min}}, P^{\mathrm{max}}$ & Grid fuse limits on aggregate power \\
        $\lambda_i$ & Reward for serving trip $i$ & $\mathcal{O} = \{(i,j) : \mathcal{T}_i \cap \mathcal{T}_j \neq \emptyset\}$ & Overlapping trip pairs \\
        \bottomrule
    \end{tabular}%
    }
\end{table}

To ensure valid operations, the system must adhere to strict physical and logical requirements:
\begin{itemize}
    \item Each trip can be served by at most one EV, and overlapping trips cannot be assigned to the same vehicle.
    \item An EV cannot be charged or discharged while actively serving a trip.
    \item The total charging and discharging power of the fleet at any time $t$ is bounded by the grid fuse constraints.
\end{itemize}

Given these settings, the objective is to maximize the number of trips served while minimizing charging costs, subject to battery capacities, grid limits, and assignment constraints. 
% EQ: Bidirectional EV charging
% EQ: Bidirectional EV charging
% EQ: Bidirectional EV charging
% EQ: Bidirectional EV charging
% EQ: Bidirectional EV charging
\begin{figure*}[t]
\centering
\begin{minipage}{0.98\textwidth}
\begin{subequations}\label{eq:bd_model_wide}
\begin{align}
\min\quad
& \Delta t \sum_{t=1}^{T}\sum_{n=1}^{N_{\mathrm{EV}}} c_t \cdot l_{n,t}
  - \sum_{i=1}^{R}\sum_{n=1}^{N_{\mathrm{EV}}} \lambda_i \cdot r_{n,i}
\label{eq:bd_obj} \\[2pt]
\text{s.t.}\quad
% & \sum_{\substack{n<m}} r_{n,i}\cdot r_{m,i} = 0, && \forall i \label{eq:bd_one_ev} \\
& \sum_n r_{n,i}\leq 1 && \forall i \label{eq:bd_one_ev} \\
% & r_{n,i}\cdot r_{n,j} = 0,
% && \forall n,\ \forall i\neq j:\ T_i\cap T_j\neq\emptyset
% \label{eq:bd_no_overlap} \\
& r_{n,i} + r_{n,j} \leq 1, 
&& \forall n, \; (i,j) \in \mathcal{O}
\label{eq:bd_no_overlap} \\
& -l_{\mathrm{max}}(1 - r_{n,i}) \leq l_{n,t} \leq l_{\mathrm{max}}(1 - r_{n,i})
&& \forall n, \; i, \; t \in \mathcal{T}_i
\label{eq:bd_no_charging} \\
& E_n^0 + \Delta t \sum_{k \leq t} l_{n,k}
  - \sum_{i:\, t_i^{\mathrm{start}}\le t} E_i^{\mathrm{trip}} \cdot r_{n,i} \ge 0,
&& \forall n,t
\label{eq:bd_soc_low} \\
& E_n^0 + \Delta t \sum_{k \leq t} l_{n,k}
  - \sum_{i:\, t_i^{\mathrm{start}}\le t} E_i^{\mathrm{trip}} \cdot r_{n,i} \le E_n^{\mathrm{cap}},
&& \forall n,t
\label{eq:bd_soc_upper} \\
& E_n^0 + \Delta t \sum_{t} l_{n,t}
  - \sum_{i} E_i^{\mathrm{trip}} \cdot r_{n,i} \ge E_n^{\mathrm{min}},
&& \forall n
\label{eq:bd_soc_final} \\
& P^{\mathrm{min}} \le \sum_{n} l_{n,t} \le P^{\mathrm{max}},
&& \forall t 
\label{eq:bd_grid_bounds} \\
& r_{n,i}\in\{0,1\}, \quad l_{n,t} \in [-l_{\mathrm{max}}, l_{\mathrm{max}}]. \notag
\end{align}
\end{subequations}
\end{minipage}
\end{figure*}
The first term in~\eqref{eq:bd_obj} represents the charging cost, where $l_{n,t}$ is the charging power of EV $n$ at time $t$. Negative levels ($l_{n,t} < 0$) indicate discharge (V2G) while positive levels ($l_{n,t} > 0$) indicate charging, and ($l_{n,t} = 0$) mean that no charging or discharge occurs at this time for that EV. The charging cost $c_{t}$ generally depends on the charging power, in particular, with different prices for selling ($l < 0$) and buying ($l > 0$) energy, but we typically use the same price for both cases, i.e., $c_{t}$ is independent of the power level. The second term of Eq.~\eqref{eq:bd_obj} rewards the trips served by setting the reward parameter $\lambda_i \geq c_{\mathrm{max}} \cdot E_{i}^{\mathrm{trip}}$. This ensures that any feasible trip is economically preferable to leaving it unassigned,
thereby implementing a ``serve-if-feasible'' logic.
Constraint~\eqref{eq:bd_one_ev} ensures that each trip is served by at most one EV.
Constraint~\eqref{eq:bd_no_overlap} prevents an EV from serving two trips with overlapping time windows. The constraint~\eqref{eq:bd_no_charging} enforces the logic that an EV cannot charge while serving. The SoC of EV must always be within the allowable range [$0, E_{n}^{\mathrm{cap}}$]. This is enforced by Eqs.~\eqref{eq:bd_soc_low} and \eqref{eq:bd_soc_upper}. 
The final SoC must be greater than $E_{n}^{\mathrm{min}}$~\eqref{eq:bd_soc_final}. 
Constraint~\eqref{eq:bd_grid_bounds} ensures that the fuse limits of the grid connection are not exceed.

The total count of SoC and grid inequalities in \eqref{eq:bd_soc_low}, \eqref{eq:bd_soc_upper}, \eqref{eq:bd_soc_final}, and \eqref{eq:bd_grid_bounds} scales as
$2N_{\mathrm{EV}}T + N_{\mathrm{EV}} + 2T$. In a standard QUBO formulation, each inequality is typically converted to an equality by introducing a slack variable. Each slack must then be discretized using binary expansion, requiring a substantial number of qubits. This overhead alone can easily exceed the practical capacity of near-term quantum hardware. To illustrate, even for a minimal instance with $(N_{\mathrm{EV}}=2, \; T = 24)$, i.e., scheduling two EVs over a single day, the model contains 146 inequalities. This motivates a decomposition-based strategy, which we develop in the following sections.
\section{Quantum-Classical Benders Decomposition}
\label{sec:hybrid}
In this section, we detail the proposed hybrid framework. Our approach is built on two foundational pillars. First, we develop a recursive SoC formulation in Sect.~\ref{sec:recursiveSoC}. 
Then, we derive logic-based conflict cuts in Sect.~\ref{sec:conflict-cuts}. 
\subsection{Classical Benders Decomposition}
\label{sec:classical-benders}
Benders decomposition is a classical technique for solving large-scale mixed-integer linear programs (MILPs).
The key idea is to identify a set of complicating variables such that, once fixed, the remaining problem reduces to a linear program that is considerably easier to solve.
The optimization then proceeds iteratively: a \emph{master problem} (MP) explores the discrete assignment space, while a continuous \emph{subproblem} (SP) evaluates the feasibility and cost of those assignments, generating cutting planes that are fed back to refine the MP~\cite{benders2005partitioning,geoffrion1972generalized}.
In the context of our bidirectional EV scheduling model, we allocate the binary trip-assignment variables~$r_{n,i}$ to the MP, while charging schedules $l_{n,t}$ and SoC dynamics are handled in the linear SP.
\subsection{Recursive SoC Formulation}
\label{sec:recursiveSoC}
The SoC bounds are the
densest part of the model. Each bound in~\eqref{eq:bd_soc_low}--\eqref{eq:bd_soc_upper} is an inequality row that references the entire charging history up to
$t$ and every trip already departed by time~$t$. We quantify the constraints density by counting non-zero elements (nnz).

For EV~$n$, the SoC bounds at time $t$ carries $t$ nnz from the charging variables. Summing this over the horizon $t=1 \dots T$ yields $T(T+1)/2$ nnz. Similarly, a trip $i$ is present from its departure onward, i.e.\ in
($T-t_i^{\mathrm{start}}+1$) rows per bound. Assuming trip start times are uniformly distributed across the horizon, a trip appears in $\mathbb{E}\left[T - t_i^{\mathrm{start}} + 1\right] = (T+1)/2$ equations on average. Across all $R$ trips, this contributes $R(T+1)/2$ nnz. 

Finally, the end-of-horizon SoC requirement~\eqref{eq:bd_soc_final} adds one row with $T$ charging nnz and $R$ assignment nnz. Summing these contributions across all $N_{\mathrm{EV}}$ vehicles, and accounting for both SoC bounds (upper and lower), the total nnz count is:
\begin{equation}
  \mathbb{E}\!\left[\mathrm{nnz}_{\mathrm{cum}}^{\mathrm{SoC+fin}}\right]
  \;=\; N_{\mathrm{EV}}\Big[\,
    \underbrace{T(T{+}1) + R(T{+}1)}_{\text{both SoC bounds}}
    \;+\; \underbrace{(T + R)}_{\text{final SoC}}\Big]
  \;=\; N_{\mathrm{EV}}\,(T+R)(T+2).
  \label{eq:nnz-cum}
\end{equation}
We introduce an explicit SoC variable $E_{n,t}$ and rewrite the battery dynamics recursively:
\begin{equation}
E_{n,t} \;=\; E_{n,t-1} \;+\; \Delta t \cdot l_{n,t}
\;-\; \sum_{i:\,t_i^{\mathrm{start}} = t} E_i^{\mathrm{trip}} \cdot r_{n,i},
\qquad \forall n, t,
\label{eq:soc-recursion}
\end{equation}
with the initial condition $E_{n,0} = E_n^0$. Observe that the dense upper and lower SoC inequalities reduce to simple variable bounds: $0 \le E_{n,t} \le E_n^{\mathrm{cap}}$. Additionally, the end-of-horizon requirement~\eqref{eq:bd_soc_final} is enforced as $E_{n,T} \ge E_n^{\min}$. This results on reduced nnz count:
\begin{equation}
\mathrm{nnz}_{\mathrm{rec}}^{\mathrm{SoC+fin}}
\;=\; N_{\mathrm{EV}}(3T + R + 1),
\label{eq:nnz-rec}
\end{equation}
which is exact for every instance. The resulting density reduction is
\begin{equation}
  \frac{\mathrm{nnz}_{\mathrm{cum}}^{\mathrm{SoC+fin}}}
       {\mathrm{nnz}_{\mathrm{rec}}^{\mathrm{SoC+fin}}}
  \;=\; \frac{(T+R)(T+2)}{3T+R+1}
  \;\xrightarrow{\;T \gg R\;}\; \frac{T}{3},
  \label{eq:nnz-ratio}
\end{equation}
i.e., the density of the SoC block drops from $\mathcal{O}(T^2)$ to $\mathcal{O}(T)$ per vehicle.

This simple reformulation has significant impact on the optimizer performance. Table~\ref{tab:nnz} compares performance on the dataset of Ref.~\cite{varga2022}, reporting the total model nnz (across all constraints) to highlight the overall sparsity gain. Observe that for the large instance $(T{=}768,\, N_{\mathrm{EV}}{=}20,\, R{=}80)$, Gurobi exhausted the memory (OOM) building the cumulative model. By contrast, the recursive formulation remains highly tractable. We adopt it hereafter for both the Benders subproblem and direct-MILP baseline.
% \begin{table*}
% \begin{table*}
\begin{table*}[t]
  \centering
  \caption{%
    Comparison of the cumulative and recursive SoC formulations. \emph{nnz}: constraint-matrix nonzero count; \emph{Speedup}: cumulative-to-recursive ratio; \emph{OOM}: Gurobi ran out of memory during model build.
  }
  \label{tab:nnz}
  \setlength{\tabcolsep}{5pt}
  \begin{tabular}{l rr rr rr}
    \toprule
    & \multicolumn{2}{c}{\textbf{Cumulative}} & \multicolumn{2}{c}{\textbf{Recursive}} & \multicolumn{2}{c}{\textbf{Speedup}} \\
    \cmidrule(lr){2-3} \cmidrule(lr){4-5} \cmidrule(l){6-7}
    \textbf{Instance} & \textbf{nnz} & \textbf{time} & \textbf{nnz} & \textbf{time} & \textbf{nnz} & \textbf{time} \\
    \midrule
    $(32,\ 1 ,\ 4)$      & 1K   & 0.01s           & 236   & 0.01s  & 5.8$\times$  & 1$\times$ \\
    $(192,\ 5 ,\ 20)$     & 220K & 2.4m           & 15K   & 15.1s & 14.4$\times$ & 9.6$\times$ \\
    $(768,\ 20 ,\ 80)$    & 14M  & \textit{OOM}   & 725K  & 20.0m & 19.1$\times$ & \textit{n/a}  \\
    \bottomrule
  \end{tabular}
\end{table*}

% \subsection{Hybrid Decomposition Architecture}
\subsection{Hybrid Decomposition Architecture}
\label{sec:hybrid-architecture}
Our goal is to decouple the NP-hard combinatorial trip assignment from the continuous energy dispatch. To achieve this, we formulate MP tailored for quantum execution. By rewriting constraints~\eqref{eq:bd_one_ev} and \eqref{eq:bd_no_overlap} in quadratic form, we eliminate the auxiliary qubits otherwise required to convert inequalities into equalities. All continuous variables and dispatch-side constraints are handled within the classical SP. The MP at iteration $k$ therefore retains only the binary variables $r_{n,i}$ and a continuous cost surrogate $\theta$:

\begin{subequations} \label{eq:MP_all}
\begin{align}
  \min_{r,\theta}\quad & \theta - \sum_{i=1}^{R}\sum_{n=1}^{N_{\mathrm{EV}}} \lambda_i\, r_{n,i}
        \tag{MP$^{(k)}$}\\
  \mathrm{s.t.}\quad & \sum_{\substack{n\neq m}} r_{n,i}\cdot r_{m,i} = 0
        && \forall i \label{eq:mp-onetrip}\\
  & r_{n,i}\cdot r_{n,j} = 0
        && \forall n,\; (i,j) \in \mathcal{O} \label{eq:mp-overlap}\\
  & \theta \ge \pi_0^{(k')} + \sum_{n,i} \pi_{n,i}^{(k')}\, r_{n,i}
        && k' \in \mathcal{K}_{\mathrm{opt}} \label{eq:mp-opt}\\
  & \sum_{i \in C_n^{(k')}} r_{n,i} \;\le\; |C_n^{(k')}| - 1
        && k' \in \mathcal{K}_{\mathrm{cnf}} \label{eq:mp-conflict}\\
  & r_{n,i} \in \{0,1\},\quad \theta \in \mathbb{R}, \nonumber
\end{align}
\end{subequations}where $\mathcal{K}_{\mathrm{opt}}$ and $\mathcal{K}_{\mathrm{cnf}}$ are the sets containing the optimality and logic-based cuts, respectively, accumulated up to iteration $k$.

For a candidate assignment $r^*$ returned by the MP, the SP is the
linear dispatch problem:
\begin{subequations} \label{eq:SP_all}
\begin{align}
\min_{l}\quad & \Delta t\, \sum_{t=1}^{T}\sum_{n=1}^{N_{\mathrm{EV}}} c_t\, l_{n,t} \tag{SP($r^*$)}\\
\mathrm{s.t.}\quad & E_{n,t} -E_{n,t-1} - \Delta t\, l_{n,t}
        = - \!\!\sum_{i:\,t_i^{\mathrm{start}}=t}\!\! E_i^{\mathrm{trip}}\, r_{n,i}^*
        && \forall n \;t 
\label{eq:sp-recursion}\\
& 0 \le E_{n,t} \le E_n^{\mathrm{cap}} && \forall n, \;t 
\label{eq:sp-soc}\\
& E_{n,T} \ge E_n^{\mathrm{min}} && \forall n 
\label{eq:sp-final_soc}\\
& -l_{\max}(1- s_{n,t}^*) \le l_{n,t} \le l_{\max}(1 - s_{n,t}^*) && \forall n, \;t
\label{eq:sp-lbnd}\\
& P_{\min} \le \sum_n l_{n,t} \le P_{\max}, && \forall t 
\label{eq:sp-fuse}
\end{align}
\end{subequations}

where $s_{n,t}^* = \min (1, \sum_{i\,:\,t \in \mathcal{T}_i} r_{n,i}^*)$
indicates whether vehicle $n$ is serving a trip at time $t$.
The interaction between these two problems is governed by Benders cuts. When SP($r^*$) is feasible, its dual variables yield an optimality cut~\eqref{eq:mp-opt}. When SP($r^*$) is infeasible, it returns feasibility cut to prevent the MP from selecting $r^*$ again. The derivation of these cuts and the
associated qubit-encoding cost are the subject of
the next subsection.

\subsection{Logic-based feasibility cuts}
\label{sec:conflict-cuts}
For some candidates $r^*$, there is no feasible $l$, and the above SP is infeasible. In standard Benders, this is handled by extracting a Farkas certificate from the dual ray of the infeasible SP and deriving a feasibility cut. However, in our V2G settings, this dual-based cut is structurally weak and exhausts many iterations. We therefore adopt logic-based cuts~\cite{hooker2003logic}. 

First, we assume that the grid power~\eqref{eq:sp-fuse} is sufficiently provisioned: $P_{\max} \ge N_{\mathrm{EV}} \cdot l_{\max}$ (e.g., this can be enforced by limiting the number of EVs entering the scheduling phase). Under this assumption, the SP is completely decoupled per EV. Therefore, an infeasible solution $r^*$ implies that at least one EV cannot commit to the assigned subset of trips $\bar{A}_n = \{i : r_{n,i}^* = 1\}$. Our task is then to identify an irreducible inconsistent subsystem (IIS) $C_n \subseteq \bar{A}_n$ that excludes this subset (and any superset of $C_n$) from the solution space.
We solve $N_{\mathrm{EV}}$ independent LPs, and for each infeasible LP we apply a deletion filter to extract the IIS $C_n$. Starting from $\bar{A}_n$, each trip is temporarily removed and per-EV LP is re-solved. If the remaining subset is still infeasible even without serving that trip, the trip is deemed non-essential to the conflict and is permanently dropped. Conversely, if removing the trip restores feasibility, the trip is essential and is retained. The resulting set $C_n$ is minimal in the sense that no proper subset of it is infeasible, and yields the combinatorial cut:\[\sum_{i \in C_n} r_{n,i} \le |C_n| - 1\]

\begin{figure}[t]
\centering
\includegraphics[width=\textwidth]{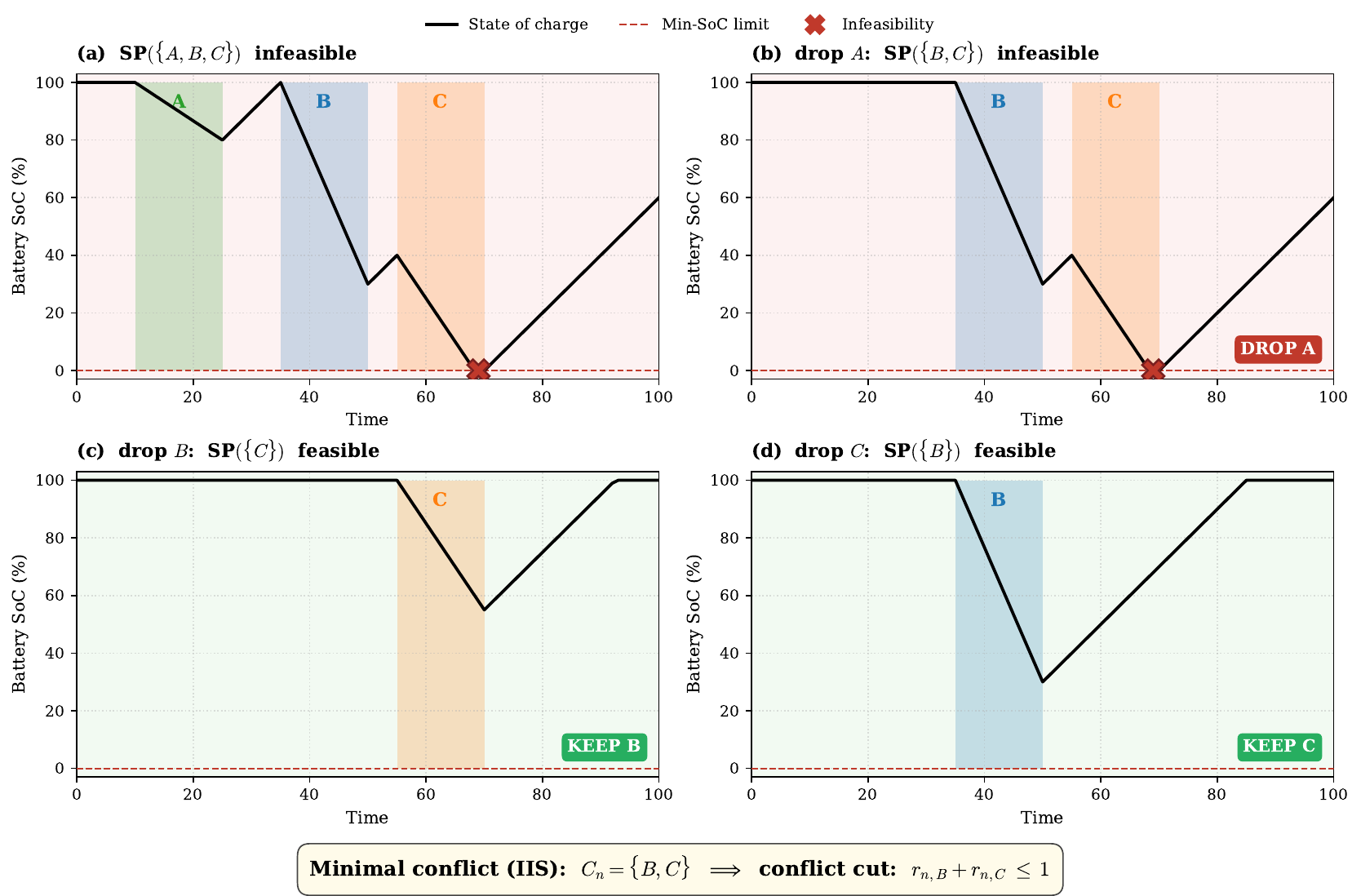}
\caption{IIS extraction illustrated on a toy instance with three trips $A, B,$ and $C$ assigned to a single EV. Panel (a) shows the candidate assignment $\bar{A}_n = \{A, B, C\}$ exhausting the battery during trip $C$. Panel (b) demonstrates that removing $A$ leaves the subset $\{B, C\}$ infeasible, identifying $A$ as non-essential to the conflict. Panels (c) and (d) verify that the remaining set $\{B, C\}$ is irreducible, as removing either element restores feasibility. The resulting minimal conflict $C_n = \{B, C\}$ generates the combinatorial cut $r_{n,B} + r_{n,C} \le 1$, strictly excluding any master assignment that allocates vehicle $n$ a superset of $\{B, C\}$. If $\bar{A}_n$ contains multiple independent conflicts, the deletion filter extracts one minimal IIS per Benders iteration, systematically eliminating the remaining conflicts in subsequent iterations.}
\label{fig:iis-extraction}
\end{figure}

% which excludes every master assignment giving vehicle $n$ any superset of $C_n$.
Empirically, we have observed that $70\text{--}85\%$ of extracted conflicts have a size $|C_n| \le 2$. This means they reduce to either single-variable fixings ($r_{n,i} = 0$) or pairwise exclusions ($r_{n,i} + r_{n,j} \le 1$). Note that we can express these pairwise cuts in a quadratic form ($r_{n,i} \cdot r_{n,j} = 0$). This has significant implications for QUBO encoding: the majority of the conflict cuts do not require auxiliary slack qubits or binary expansions to encode. We revisit this in Sec.~\ref{sec:results} and report per-instance auxiliary-qubit budgets. The IIS extraction procedure is illustrated in Fig.~\ref{fig:iis-extraction}, and the hybrid framework is given in Algorithm~\ref{alg:hbd}.
% \begin{algorithm}[t]
% \begin{algorithm}[t]
\begin{algorithm}[t]
\caption{Hybrid Quantum-Classical Benders Decomposition}
\label{alg:hbd}
\begin{algorithmic}[1]
\State Initialise MP with structural constraints~\eqref{eq:mp-onetrip}-\eqref{eq:mp-overlap} and $\bar z \gets +\infty$,\; $\underline z \gets -\infty$
\State Add pre-seed optimality cut from SP($\bar r = 0$)
\Repeat
  \State Solve MP on quantum hardware $\rightarrow$ candidate $\bar r$, lower bound $\underline z$
  \State Solve SP($\bar r$) classically
  \If{SP feasible}
   \State $z_{\mathrm{cand}} \gets \mathrm{SP}(\bar r) - \sum_{n,i}\beta_i\, \bar r_{n,i}$;\;
          if $z_{\mathrm{cand}} < \bar z$, set $\bar z \gets z_{\mathrm{cand}}$ and store incumbent
   \State Add optimality cut~\eqref{eq:mp-opt} from SP duals
\Else
     \For{each infeasible vehicle $n$}
        \State $C_n \gets \textsc{IIS-extract}(\bar{A}_n)$
              \Comment{$\bar{A}_n = \{i : \bar r_{n,i} = 1\}$; deletion filter}
        \State Add conflict cut~\eqref{eq:mp-conflict}
     \EndFor
  \EndIf
\Until{$(\bar z - \underline z) / |\bar z| < \varepsilon$ or iteration limit}
\State \Return $(r^\star, l^\star, E^\star)$
\end{algorithmic}
\end{algorithm}

% figures/iis_extraction_panels_v2.pdf
% figures/iis_extraction_panels_v2.pdf
% figures/iis_extraction_panels_v2.pdf
% figures/iis_extraction_panels_v2.pdf

% \subsectionQuantum Formulation}
% \subsectionQuantum Formulation}
\subsection{MP Quantum Formulation}
We solve the MP using the QAOA algorithm~\cite{farhi2014}.
For this, we first transform the MP to its quantum formulation by selecting a computation basis and  replacing the search variables with the corresponding quantum operators.
Then we incorporate all constraints as penalty terms. The computational basis states are indexed by the qubit variables $r_{n,i} \in \{0,1\}$. The corresponding $z$-angular momentum operators are defined as:
\begin{align*}
R^z_{n,i} \ket{r_{n,i}} &= r_{n,i}\ket{r_{n,i}}, \ \\
R^{+}_{n,i} \ket{r_{n,i}} &= \sqrt{(1-r_{n,i})(1+r_{n,i})}\ket{r_{n,i}+1}, \ \\
R^{-}_{n,i} \ket{r_{n,i}} &= \sqrt{r_{n,i},(2-r_{n,i})} \ket{r_{n,i}-1}.
\end{align*}
The $x$ and $y$ operators are defined as $R^x_{n,i} = \tfrac{1}{2}(R^+_{n,i}+R^-_{n,i})$ and $R^y_{n,i} = \tfrac{1}{2i}(R^+_{n,i}-R^-_{n,i})$. 
The assignment register spans a Hilbert
space of dimension $2^{N_{\mathrm{EV}}R}$. The encoding additionally allocates a
binary-expanded register for the surrogate variable $\Theta$.
The MP cost Hamiltonian at
iteration $k$ is
\begin{align}
H_{\mathrm{MP}}^{(k)} \;=\;&
  \Theta \;-\; \sum_{i=1}^{R}\sum_{n=1}^{N_{\mathrm{EV}}} \lambda_i\, R^z_{n,i}
  && \text{(trip rewards)}\nonumber\\
& +\; P \sum_{i=1}^{R}\sum_{n<m} R^z_{n,i}\, R^z_{m,i}
  && \text{(at most one EV per trip)}\nonumber\\
& +\; P \sum_{n=1}^{N_{\mathrm{EV}}}
       \sum_{(i,j)\in\mathcal{O}} R^z_{n,i}\, R^z_{n,j}
  && \text{(no overlapping trips)}\nonumber\\
& +\; P \!\!\!\sum_{k'\in\mathcal{K}_{\mathrm{opt}}}\!\!\! H^{(k')}_{\mathrm{opt}}
  \;+\; P \!\!\!\sum_{k'\in\mathcal{K}_{\mathrm{cnf}}}\!\!\! H^{(k')}_{\mathrm{cnf}},
\end{align}
where $P$ is a penalty factor. Each optimality cut \eqref{eq:mp-opt} is converted to an equality by introducing a non-negative slack variable $S^{(k')}_{\mathrm{opt}}$:\begin{equation}
H^{(k')}_{\mathrm{opt}}
\;=\; \Bigl(\,\Theta - \pi_0^{(k')}
       - \sum_{n,i}\pi^{(k')}_{n,i}\, R^z_{n,i}
       - S^{(k')}_{\mathrm{opt}}\Bigr)^{\!2}.
\label{eq:H-opt}
\end{equation}
Similarly, conflict cuts~\eqref{eq:mp-conflict} are enforced independently for each EV:
\begin{equation}
H^{(k')}_{\mathrm{cnf}}
\;=\; \Bigl(\,\sum_{i \in C^{(k')}_n} R^z_{n,i}
       + S^{(k')}_{\mathrm{cnf}} - |C^{(k')}_n| + 1\Bigr)^{\!2},
\label{eq:H-cnf}
\end{equation}
with integer slack $S^{(k')}_{\mathrm{cnf}}\in\{0,\dots,|C^{(k')}_n|-1\}$. For the dominant
case $|C^{(k')}_n|=2$ (Sect.~\ref{sec:conflict-cuts}) this collapses to the slack-free
pairwise penalty $P\,R^z_{n,i}\,R^z_{n,j}$, requiring no auxiliary qubits.
\section{Experimental Results}
\label{sec:results}

% Fig.3 conflict cuts distribution
% Fig.3 conflict cuts distribution
\begin{figure}[t]
    \centering
    \includegraphics[width=1\linewidth]{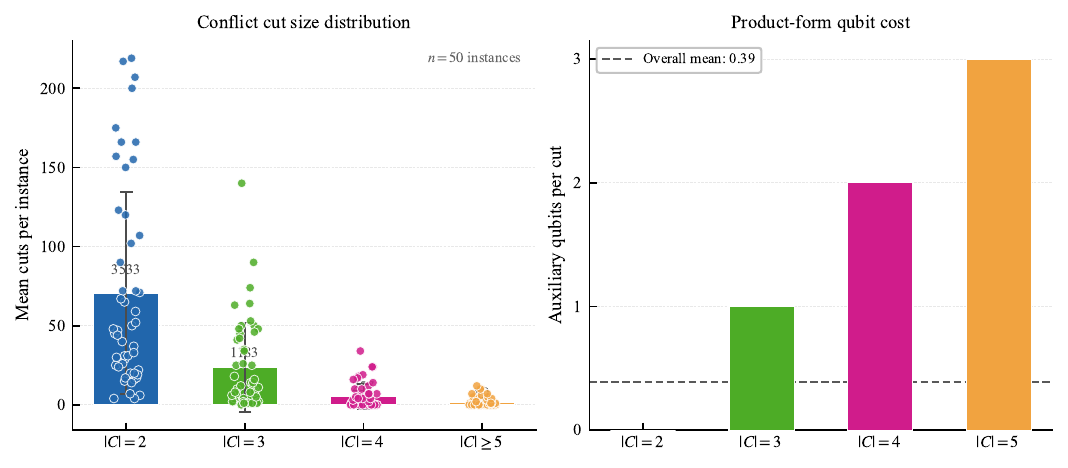}
    \caption{Distribution of logic-based conflict cut sizes (left) and their associated QUBO encoding costs (right) across 50 medium instances.}
    \label{fig:cut-distribution}
\end{figure}

% Table 3 optimization results
% Table 3 optimization results
\begin{table}[t]
  \centering
  \setlength{\tabcolsep}{5pt}
  \caption{Per-instance results on medium instances ($T=192$, $N=5$, $R=20$). Benders convergence gap $(UB-LB)/|UB| \times 100$ at termination, and Relat.\ gap $=$ $(z^*_{\text{B}} - z^*_{\text{dir}}) / |z^*_{\text{dir}}| \times 100$.}
  \label{tab:medium_results}
  \begin{tabular}{l rr@{\hspace{6pt}} rrrrrr r} 
    \toprule
    & \multicolumn{2}{c}{Direct MILP}
    & \multicolumn{7}{c}{Hybrid Benders} \\ 
    \cmidrule(lr){2-3}\cmidrule(l){4-10} 
    Inst & Obj.\ & Time & Obj.\ & Time & Iters & \makecell{Conf.\\ cuts} & \makecell{Aux\\qubits} & \makecell{Bdrs\\gap (\%)} & \makecell{Relat.\\gap (\%)}\\
    \midrule
     01 &  -7,491.01 &   62 &  -7,487.82 &  101 &    31 &    22 &      1 &   1.00 & 0.04\\
     02 &  -9,506.15 &  125 &  -9,503.36 &   75 &    33 &    64 &     14 &   0.94 & 0.03\\
     03 &  -9,155.98 &  324 &  -9,149.74 &  600 &    42 &    32 &      6 &   0.99 & 0.07\\
     04 &  -9,536.58 &  196 &  -9,529.30 &   83 &    35 &    38 &      1 &   0.96 & 0.08\\
     05 &  -7,867.73 &   76 &  -7,864.21 &   50 &    29 &    24 &      5 &   0.98 & 0.04\\
    \midrule
    % 4. Added the final value for the new column (replace 0.05 with your actual mean!)
    \textit{Mean} &  -8,711.49 &  156.6 &  -8,706.88 &  181.9 &    34 &    36 &    5.4 &   0.97 & 0.05 \\
    \bottomrule
  \end{tabular}
\end{table}

For benchmarking, we use the dataset of Ref.~\cite{varga2022}, which provides 810 instances of varying sizes and configurations. We adapt the dataset to our bidirectional use case by replacing the non-negativity bound on $l_{n,t}$ with a symmetric range $-l_{\max} \le l_{n,t} \le l_{\max}$. The remaining parameters are set as follows:
\begin{itemize}
  \item Trip-service reward: $\lambda_i = 2\, c_{\max}\, E_i^{\mathrm{trip}}$
  \item Final-SoC requirement: $E_n^{\mathrm{min}} = 0.1 \cdot E_n^{\mathrm{cap}}$
  \item Fuse limits: $P_{\max} = -P_{\min} = N_{\mathrm{EV}} \cdot l_{\max}$.
\end{itemize}

The classical computations were performed on a Dell laptop with an Intel Core i7-1185G7 processor (4 cores, 8 logical threads) and 16 GB RAM.

First, we demonstrate the effectiveness of our logic-based cuts. Fig.~\ref{fig:cut-distribution} presents the distribution of conflict cut sizes across 50 medium instances. The left panel shows that the majority of cuts are size $|C_n| = 2$, while cuts with $|C_n| \geq 5$ are rare and cluster near
zero. We also observed cuts of size $|C_n| = 1$. These arise when a single trip is infeasible on its own (e.g., when a trip starts so early in the horizon that the vehicle's initial SoC, $E_n^0$, cannot be raised to the required $E_i^{\mathrm{trip}}$ before $t_i^{\mathrm{start}}$). The right panel illustrates the auxiliary qubit cost required for each cut size. Crucially, size-2 cuts require zero additional qubits for their QUBO encoding.
In Table~\ref{tab:medium_results}, we report the optimization results of 5 medium-sized instances. 
Note, that these instances are already far too large for pure quantum approaches~\cite{ekstrom2026qudit}.  
Overall, the hybrid approach terminates at a Benders convergence tolerance of $\le 1\%$ on all instances, and successfully recovers the MILP optimum with a mean relative gap of $0.05\%$. For some instances, the computational time of the hybrid decomposition is larger than that of the direct solution, an expected result, since Benders advantages become more pronounced for large-scale problems. Interestingly, the qubit overhead for encoding conflict cuts is rather small. The largest count is 14 which is needed to encode 64 total cuts. Instances 1 and 4 require only 1 qubit for encoding their total cuts, which implies only 1 cut has size $>2$.

As a proof of concept, we also executed a small-scale instance ($T=3, N_{\mathrm{EV}}=2, R=2$) on real IBM quantum hardware. At each Benders iteration, the MP is solved on the IBM backend using QAOA. From the sampled bitstring distribution, we extract the top-$k$ most frequent candidates and evaluate them in batch using the SP feasibility checks. For this reduced instance, the method converges within a single Benders iteration and returns the same optimal solution as the classical baseline. Table~\ref{tab:benders_vs_milp} summarizes this convergence against the Gurobi MILP solver.

\begin{table}[t]
\centering
\begin{tabular}{lccccr}
\toprule
\textbf{Method} & \textbf{Iteration} & \textbf{Lower Bound} & \textbf{Upper Bound} & \textbf{Gap} & \textbf{Cuts Added} \\
\midrule
Hybrid Benders  & 0 & $-10000$ & $10000$ & $20000$ & 4 \\
                & 1 & $-350$   & $-350$  & $0$     & -- \\
\addlinespace
Gurobi MILP     & -- & $-350$  & $-350$  & $0$     & -- \\
\bottomrule
\end{tabular}
\caption{Comparison of the hybrid Benders approach (via QAOA) and the classical MILP baseline on a small-scale instance ($T=3, N_{\mathrm{EV}}=2, R=2$).}
\label{tab:benders_vs_milp}
\end{table}

% \section{Conclusion}
\section{Conclusion}
In this paper, we have presented a hybrid quantum-classical framework for the joint optimization of large-scale bidirectional EV charging and trip assignment problems. By adopting a recursive SoC formulation, we improved the sparsity of the constraint matrix and consequently reduced memory and time requirements. Additionally, the introduction of conflict cuts to handle Benders feasibility cuts resulted in an efficient MP formulation with near-zero-qubit encoding overhead. The framework recovers MILP-optimal objectives on the benchmark instances within a small auxiliary-qubit budget, with end-to-end execution validated on IBM quantum hardware. Although our conflict cuts contribute significantly to the overall reduction of the qubit encoding cost, the total qubit budget remains dominated by Benders optimality cuts. Moving forward, we plan to address these optimality cuts and investigate cut-strengthening techniques.
\section*{Acknowledgements}
Funded by the European Union under Horizon Europe Programme - Grant Agreement 101080086 — NeQST.\\
Views and opinions expressed are however those of the author(s) only and do not necessarily reflect those of the European Union or European Climate, Infrastructure and Environment Executive Agency (CINEA). Neither the European Union nor the granting authority can be held responsible for them.

\printbibliography

@techreport{iea2025,
  author      = {{International Energy Agency}},
  title       = {Global {EV} Outlook 2025},
  institution = {IEA},
  year        = {2025},
  note        = {Available: \url{https://www.iea.org/reports/global-ev-outlook-2025}}
}

@article{kumar2025comprehensive,
  title={A comprehensive review of vehicle-to-grid integration in electric vehicles: Powering the future},
  author={Kumar, Pulkit and Channi, Harpreet Kaur and Kumar, Raman and Rajiv, Asha and Kumari, Bharti and Singh, Gurpartap and Singh, Sehijpal and Dyab, Issa Farhan and Lozanovi{\'c}, Jasmina},
  journal={Energy Conversion and Management: X},
  volume={25},
  pages={100864},
  year={2025},
  publisher={Elsevier}
}

@article{golsefidi2025,
  author  = {Golsefidi, A.~H. and Hipolito, F. and others},
  title   = {Incremental expansion of large scale fixed and mobile charging
             infrastructure in stochastic environments: A novel graph-based
             {Benders} decomposition approach},
  journal = {Appl.\ Energy},
  volume  = {380},
  pages   = {124985},
  year    = {2025},
  doi     = {10.1016/j.apenergy.2024.124985}
}

@article{arslan2016,
  author  = {Arslan, O. and Kara\c{s}an, O.~E.},
  title   = {A {Benders} decomposition approach for the charging station
             location problem with plug-in hybrid electric vehicles},
  journal = {Transp.\ Res.\ B Methodol.},
  volume  = {93},
  pages   = {670--695},
  year    = {2016},
  doi     = {10.1016/j.trb.2016.09.004}
}

@article{ccalik2019benders,
  title={A Benders decomposition method for locating stations in a one-way electric car sharing system under demand uncertainty},
  author={{\c{C}}al{\i}k, Hatice and Fortz, Bernard},
  journal={Transportation Research Part B: Methodological},
  volume={125},
  pages={121--150},
  year={2019},
  publisher={Elsevier}
}

@article{ren2025electric,
  title={The electric vehicle routing problem of a new mobile charging service},
  author={Ren, Kanghui and Li, Maosheng and Cen, Xuekai and Huang, Helai},
  journal={Transportation Safety and Environment},
  volume={7},
  number={1},
  pages={tdae012},
  year={2025},
  publisher={Oxford University Press}
}

@misc{farhi2014,
  author        = {Farhi, E. and Goldstone, J. and Gutmann, S.},
  title         = {A quantum approximate optimization algorithm},
  year          = {2014},
  eprint        = {1411.4028},
  archivePrefix = {arXiv},
  primaryClass  = {quant-ph}
}

@article{quantum_cslp2024,
  author  = {Rao, P.~U. and Sodhi, B.},
  title   = {Hybrid quantum-classical solution for electric vehicle
             charger placement problem},
  journal = {Soft Comput.},
  volume  = {27},
  number  = {18},
  pages   = {13003--13016},
  year    = {2023},
  doi     = {10.1007/s00500-023-08876-5}
}

@article{mussig2023solving,
  title={Solving the Optimization Problem of Bidirectional EV-Charging on a Quantum Computer},
  author={M{\"u}ssig, Daniel and Federer, Marika and Lenk, Steve and L{\"a}ssig, J{\"o}rg},
  journal={Harzer Hochschultexte, 14th ser. NWK},
  pages={94--101},
  year={2023}
}

@inproceedings{veshchezerova2023hybrid,
  title={A hybrid quantum-classical approach to the electric mobility problem},
  author={Veshchezerova, Margarita and Somov, Mikhail and Bertsche, David and Limmer, Steffen and Schmitt, Sebastian and Perelshtein, Michael and Tripathi, Ayush Joshi},
  booktitle={2023 IEEE International Conference on Quantum Computing and Engineering (QCE)},
  volume={1},
  pages={636--641},
  year={2023},
  organization={IEEE}
}

@article{du2025smart,
  title={Smart ride and delivery services with electric vehicles: Leveraging bidirectional charging for profit optimisation},
  author={Du, Jinchun and Shen, Bojie and Cheema, Muhammad Aamir and Toosi, Adel N},
  journal={Information Sciences},
  pages={122929},
  year={2025},
  publisher={Elsevier}
}

@article{varga2022,
  author  = {Varga, J. and Raidl, G.~R. and Limmer, S.},
  title   = {Computational methods for scheduling the charging and
             assignment of an on-site shared electric vehicle fleet},
  journal = {IEEE Access},
  volume  = {10},
  pages   = {105786--105806},
  year    = {2022},
  doi     = {10.1109/ACCESS.2022.3210168}
}

@article{benders2005partitioning,
  author  = {Benders, J.~F.},
  title   = {Partitioning procedures for solving mixed-variables
             programming problems},
  journal = {Comput.\ Manag.\ Sci.},
  volume  = {2},
  number  = {1},
  pages   = {3--19},
  year    = {2005},
  doi     = {10.1007/s10287-004-0020-y},
  note    = {Reprinted from Numerische Mathematik, 1962}
}

@article{geoffrion1972generalized,
  title={Generalized benders decomposition},
  author={Geoffrion, Arthur M},
  journal={Journal of optimization theory and applications},
  volume={10},
  number={4},
  pages={237--260},
  year={1972},
  publisher={Springer}
}

@article{hooker2003logic,
  author  = {Hooker, J.~N. and Ottosson, G.},
  title   = {Logic-based {Benders} decomposition},
  journal = {Math.\ Program.},
  volume  = {96},
  number  = {1},
  pages   = {33--60},
  year    = {2003},
  doi     = {10.1007/s10107-003-0375-9}
}

@article{deller2023quantum,
  title={Quantum approximate optimization algorithm for qudit systems},
  author={Deller, Yannick and Schmitt, Sebastian and Lewenstein, Maciej and Lenk, Steve and Federer, Marika and Jendrzejewski, Fred and Hauke, Philipp and Kasper, Valentin},
  journal={Physical Review A},
  volume={107},
  number={6},
  pages={062410},
  year={2023},
  publisher={APS},
  doi={10.1103/PhysRevA.107.062410}
}

@article{ekstrom2026qudit,
  title={Comparing Qubit and Qudit Encodings for EV Charging and Trip Assignment Problems},
  author={Linus Ekstrøm and Hao Wang and Sebastian Schmitt},
  note={Accepted at GECCO 2026},
  journal={arXiv:2605.10255},
  year={2026},
  doi={10.48550/arXiv.2605.10255}
}
\end{document}